\documentstyle[aps,prl,twocolumn]{revtex}

\begin{document}
\author{V. V. Makhro\footnote{E-mail: maxpo@chat.ru}}
\address{Bratsk State Industrial University, Bratsk, 665709, Russian Federation}
\title{Tunneling in the moving domain wall}
\maketitle

\begin{abstract}
The mechanism of the parametrical stimulated tunneling in the
spectrum of the moving domain wall considered. It is shown that
such a mechanism can to cause the initial phase of the
parametrical evolution of domain wall's surface waves. In turn,
this evolution leads to an appearance of essential features in
domain walls dynamics.
\end{abstract}

In a middle 80th a new magnetic materials with an unique physical
and technological properties became available for the
investigations. It is so-called orthoferritin. By its magnetic
nature orthoferritin \cite{1} are the weak ferromagnets and may be
considered within a framework of a double-sublattice model by
introducing the vectors of ferromagnetism ${\bf m}
$ and antiferromagnetism ${\bf l}$ so that ${\bf lm}=0$ and $%
l^{2}=1-m^{2}\approx 1$. In spherical coordinates ($l_{x}=\sin \theta \cos
\varphi$, $l_{y}=\sin \theta \sin \varphi$, $l_{z}=\cos \theta$) the
Lagrangian density will be given by the expression
\begin{equation}
{\cal L}=\frac{\chi_{\perp}}{2\gamma ^{2}}(\dot{{\bf l}})^{2}-\frac{\chi
_{\perp}}{\gamma}{\bf H}{\bf [l\dot{l}]}-\Phi,
\end{equation}
where the thermodynamical potential $\Phi$ is \cite{Belov}
\[
\Phi = A(\nabla {\bf l})^{2}-\frac{\chi _{\perp}}{2}\left(H^{2}-({\bf Hl}%
)^{2}\right)-M_{z}^{0}H_{z}l_{x}-M_{x}^{0}H_{x}l_{z}+
\]
\begin{equation}
+K_{ac}l_{z}^{2}-K_{ab}l_{x}^{2},
\end{equation}
and $A$ - uniform exchange constant, $K_{ac}$ and $K_{ab}$ - anisotropy
constants, ${\bf H}$ - total external field, $\chi_{\perp}$ - transverse
susceptibility, $M_{x}^{0}$ and $M_{z}^{0}$ are the values of magnetization
in the phases $\Gamma _{2}(F_{x}C_{y}G_{z})$ and $\Gamma
_{4}(G_{x}A_{y}F_{z})$, respectively. Corresponding Lagrange equation for
the domain wall moving in a weak ferromagnet may be written as
\[
\theta^{\prime\prime}_{\xi \xi} = -\sin \theta \cos \theta,
\]
where $\xi =\frac{x-vt}{\Delta}$, $\Delta = \Delta _{0}\sqrt{1-\frac{v^{2}}{%
c^{2}}}$, and $\Delta _{0}= \sqrt{\frac{A}{K_{ac}+\chi _{\perp}H^{2}}}$ is a
width of the resting wall. By $v$ we denote the velocity of the wall, and $%
c=\gamma \sqrt{A/\chi_{\perp}}$ is a limiting velocity for the domain walls
motion, which coincides with a spin wave velocity.

One of the major circumstances which made the orthoferritin so
unique, is an extremely high value of the limiting speed $c$,
which a several times as much the velocity of sound in such a
materials ($c=2\cdot 10^{6}$ $cm/s$). It makes possible very
exotic processes of the interaction of domain walls with
environment such as Cherenkov's phonon radiation or the surface
magnons emission. Recently last phenomena attract again the
attention of the experimentalist due to improvement on
experimental technique \cite{Chetkin98}.

The early investigations in 80th reveals a very interesting features of the
domain walls dynamics in yttrium ortoferritin and iron borate: an anomalous
behavior of domain walls mobility was discovered. Domain walls velocity
growth up monotonically with increasing of the external field in regular
case. But for the some intervals of the field values the saturation of the
mobility has been occurred: there was no increasing of velocity with a
growth of a field. At the same time, later on, for the biggest magnitude of
the field the increasing of velocity recommenced. Such a behavior leads to
occurrence of the plateaus on the plots of $v(H)$.

Some of such mobility anomalies may be easily interpreted as a result of
Cherenkov's processes: the emission of magnons and phonons with a different
polarizations by the wall, because of the plateaus has been occurred for the
values of walls velocities which coincides with corresponding velocities of
the spin or elastic waves \cite{1}. But for the explanation of another
finding in experiments similar plateaus there were no appropriate
Cherenkov's processes and one must looking for another physical reasons,
which may leads to occurrence of a such anomalous.

A suitable physical mechanism has been proposed by Zvezdin and
Popkov \cite {Popkov84}, and Makhro and Kazakov \cite{Makhro84}.
Such a mechanism may be described as follows. Let's domain wall
will be moving in the periodically non-uniform magnetic media.
Further, let's some internal degrees of freedom are intrinsical
for the domain wall, and we can to characterize theirs by some
frequencies $\omega _{i}$. These degrees of freedom will excites
every time when the frequency of the perturbation arising due to
moving through non-uniform media will coincides with one of an
eigen frequencies $\omega_{i}$: $\omega_{i}=2\pi u_{i}/a$, where
$a$ is a space period of the non-uniformity. In that way the
outflow of the energy of the external field on the internal
degrees of freedom will arise when wall will arrive at
``resonant'' velocities $u_{i}$. As a consequence for such a
velocities deceleration of the wall will occur.

Lately it became clear that interaction of the wall with media
non-uniformities has a parametrical character \cite{Makhro86}, \cite
{Makhro87}: the effective elasticity of the wall which moving in
periodically non-uniform media may be considered as a function of a time and
a velocity of motion.

Indeed, if one use for the description of domain wall in the weak
ferromagnet the model of a flat membrane, than in the terms of the so-called
``reduced'' description (see Ref. \cite{Zvezdin79}) one may write for the
domain wall in periodically non-uniform media the equation of motion as
\begin{equation}
\frac{\partial}{\partial t}(m\dot{x})+m\dot{x}\Gamma-\nabla _{\perp}\sigma
\nabla _{\perp}x=2M_{0}H+\frac{U_{0}}{a}\sin \frac{2\pi x}{a},
\end{equation}
where $x$ - is a coordinate of a center of a wall, $m$ - the
effective mass of the wall\footnote{Here and below we give all
quantities per unit area of the domain wall.}, $\Gamma$ -
phenomenological constant of a dissipation and $\sigma=mc^{2}$.

Let's suppose that the wall moving with a constant speed $\dot{x}\equiv u$
and then let's denote by $q$ deflection of the points of a walls surface
from the equilibrium position. Then for the $x=ut+q$ one can write
\begin{equation}
m_{\perp }\ddot{q}+m_{\perp }\dot{q}\Gamma -m_{\parallel }c^{2}\nabla
_{\perp }^{2}q=-\frac{4\pi ^{2}U_{0}}{a^{2}}q\cos \frac{2\pi ut}{a},
\end{equation}
\[
m_{\perp }=m_{0}(1-u^{2}/c^{2})^{-3/2}),m_{\parallel }=m_{0}(1-u^{2}/c^{2}).
\]
The solution of Eq. (4) will be
\begin{equation}
q=q_{0}\exp \left( -\frac{\Gamma t}{2}-i{\bf k}_{\perp }{\bf r}_{\perp
}\right) \Phi \left( \frac{\pi ut}{a}\right) ,
\end{equation}
where $\Phi \left( \frac{\pi ut}{a}\right) $ is a solution of the Mattiew
equation
\begin{equation}
\ddot{\Phi}\left(\frac{\pi ut}{a}\right)+\left(\lambda +b\cos
\frac{2\pi ut}{a}\right)\Phi\left(\frac{\pi ut}{a}\right)=0,
\end{equation}
with a parameters
\[
\lambda=\frac{m_{\parallel}c^{2}k^{2}a^{2}}{\pi
^{2}m_{\perp}u^{2}}-\frac{\Gamma ^{2}a^{2}}{4\pi ^{2}u^{2}}\equiv
\frac{a^{2}[(c^{2}-u^{2})k^{2}-\Gamma ^{2}/4]}{\pi ^{2}u^{2}},
\]
\[
b=2U_{0}/m_{\perp}u^{2}.
\]

Formally, one can say that for the velocities which determined by
the condition $\lambda = n^{2}$ ($n$ - integer) the parametrical
resonance of the surface vibrations will occur and the velocity of
the wall will decrease. The values of a such resonant velocities
are given by the expression
\begin{equation}
u_{n}=\sqrt{\frac{a^{2}c^{2}k^{2}-a^{2}\Gamma ^{2}/4}{\pi
^{2}n^{2}+a^{2}k^{2}}}.
\end{equation}
Let's note, however, that the Eq. (4) is not a self-consistent
equation. Indeed, for any nonzero value of $\Gamma$ in resonant
case the velocity doesn't stabilezed to the one of $u_{n}$ but
become to decrease. In this reason the parameters of (6) go off
the instability region and parametrical pumping into internal
degrees of freedom come to the end. In principle, the problem
required the solving of the system of coupled equations as follows
\begin{equation}
\ddot{x}+\Gamma
\dot{x}-\frac{1}{m_{0}}\left[2M_{0}H-F_{T}+\frac{U_{0}}{a}\sin
\frac{2\pi
x}{a}\left(1-\frac{\dot{x}^{2}}{c^{2}}\right)^{1/2}\right]=0,
\end{equation}
\begin{equation}
\ddot{q}+\Gamma \dot{q}-c^{2}\nabla ^{2}_{\perp}
q\left(1-\frac{\dot{x}^{2}}{c^{2}}\right)+\frac{4\pi
^{2}qU_{0}}{m_{0}a} \cos \frac{2\pi
\dot{x}t}{a}\left(1-\frac{\dot{x}^{2}}{c^{2}}\right)=0,
\end{equation}
where $F(T)$ is a braking force, which can be calculated as
$F_{T}=\frac{d}{dx}\sum _{\mathbf{k}}m\omega
_{\mathbf{k}}^{2}q_{\mathbf{k}}^{2}.$ Numerical solving of the
system (8)-(9) has been firstly carried out in Ref.
\cite{Makhro87}. The results of \cite{Makhro87}(especially, a
width and disposal of plateau)  demonstrate a good qualitatively
agreement with experiments.

But the next circumstance remained not clear enough: parametrical
resonance for its beginnings required of existence of the initial
``inoculating'' flexure $q_{0}$, without such a flexure the
resonance can't appear. It is clear that script of parametrical
evolution of surface vibrations is very sensitive to the magnitude
of the initial flexure. In Ref. \cite{Makhro87} it was supposed
that arising of the initial flexure is conditioned by the
interaction of the wall with a non-regular magnetic defects of the
media or by the random fluctuation of the external field. But it
is difficult to agree with such an explanations. Indeed, in
experiments the plateaus on the dependence $v(H)$ has been
observed always for the same velocities, it will be impossible if
the initial flexure has a random value of a magnitude: the
evolution of the surface vibrations must follows to the different
scripts for the each time. Thus, the problem of explanation of the
regularity of the plateaus disposal now as before is actual.

We propose to discuss the next mechanism which can not only to
explain the initial flexure origin but also to be able to describe
some another features in high-speed domain walls dynamics. This is
a resonant tunneling in the spectrum of excitations of a weak
ferromagnet. In such a spectrum the ground state, which
corresponds to natural oscillations of the wall as a whole, is
separated from the excited states which corresponds to states,
with a surface (or Winter's, \cite{Winter})magnons, by the energy
gap with a width $\sigma$. The transition from the ground state
into the first excited state just corresponds to the occurrence of
the initial flexure. However, the width of a gap in orthoferritin
as a rule is large enough and both tunneling and thermal
activation has a neglected probability even for the room
temperatures. Nevertheless, some mechanism exists which can to
increase this probability and made it more or less appreciable.
Such mechanism is a resonant stimulation of tunneling. For the
first time it was described by Lin and Ballentine \cite{Valentin}
who found by numerical simulation that a monochromatic external
field acting on a quartic double-well oscillator can increase the
rate of coherent tunneling to values several orders of magnitude
higher than those for the undriven system. Later, in Ref.
\cite{Makhro98} the similar problem of the thermal resonant
stimulation of the the tunneling has been considered and it was
shown that the frequency of stimulating factor must be close to
the eigenfrequency of the inverted barrier's potential. We propose
to apply such an approach for the solving of our problem too.

Let's map the initial problem of the tunneling between ground and
first excited states onto particle problem \cite{Scharf}: such a
double-level problem can be reduced to problem of the tunneling in
the double-well potential
\begin{equation}
V=\sigma(x^{2}-\eta ^{2})^{2}.
\end{equation}
If one take into account the usual for the orthoferritin values
$K_{ac}=1\cdot 10^{5}$ $erg/cm^{3}$, $A=1\cdot 10^{-7}$ $erg/cm$,
$M^{0}_{z}= 10$ $emu$, $c=2\cdot 10^{6}$ $cm/s$, then $\sigma _{0}
=mc^{2}=4\sqrt{AK_{ac}}=0.4$ $erg/cm^{2}$, and if one accept the
usual for the real experiments (see, for example Ref.
\cite{Chetkin98}) values of the walls square $S\sim 10^{-8}$
$cm^{2}$, then for the  $\sigma$ one obtain the value $10^{-15}$
$erg$. Such a values as it was mentioned above gives for both the
usual tunneling rate and probability of the thermal activation the
neglected values.

We shall presume here as in Ref. \cite{Makhro86} that parametrical
disturbance will be acted on the domain wall moving through a
periodical magnetic non-uniform media because of its
``elasticity'' (and, as consequence $m$ and $\sigma$) becomes a
periodical functions of the time and the velocity of moving.

Let's, for the concreteness, the ground state corresponds the
localization of the particle in the left well of the effective
potential. The wave function in such a case one can close
approximate by the wave function of the ground state of the
harmonic oscillator with the eigenfrequency $\omega
_{0}=\sqrt{V''(x_{min}=\eta)/m}=\sqrt{8\sigma \eta ^{2}/m}$
\cite{Pipa}
\begin{equation}
\Psi _{0}=e^{-1/2(\xi ^{2}+i\tau)},
\end{equation}
where
\[
\xi =x\sqrt{\frac{m\omega _{0}}{\hbar}},\tau=\omega _{0}t.
\]
The evolution of Eq. (11) can be described by
\begin{equation}
\frac{\partial ^{2}\Psi}{\partial \xi ^{2}}-\xi ^{2}(1-\beta \cos
2r\tau)\Psi=-2i \frac{\partial \Psi}{\partial \tau},
\end{equation}
where $\beta$ is  coefficient of parametrical modulation, and $r$
is a frequency of parametrical perturbation (with the assumption
of the fine adjustment on the main resonance $r=\pi u/a$). The
solution of Eq. (12) one can found as
\begin{equation}
\Psi=e^{-y\xi ^{2}+k}.
\end{equation}
Eq. (12) has a sense  when the next conditions will be fulfilled
\begin{equation}
k=i\int_{0}^{\tau}yd\tau,
\end{equation}
and
\begin{equation}
2i\frac{dy}{d\tau}-4y^{2}+1+\beta \cos {2rt}=0.
\end{equation}
By the introduction of the function $u(\tau)$:
\[
2iuy=\frac{du}{d\tau},
\]
Eq. (15) may be reduced to the canonical form
\begin{equation}
\frac{d^{2}u}{d\tau ^{2}}+(1+\beta \cos 2rt)u=0.
\end{equation}
In the case of the fine resonance the solution of Eq. (16) is
\begin{equation}
u=e^{\Delta \tau} \cos (\tau+\frac{1}{4} \pi)+Ce^{-\Delta
\tau}\cos(\tau-\frac{1}{4}\pi),
\end{equation}
where $\Delta=\frac{1}{4}\beta$ and $C$ is a constant of
integration.

After returning to $y$ one can obtain
\begin{equation}
y(\tau)=\frac{i}{2}\frac{e^{\Delta \tau}\sin (\tau
+\frac{1}{4}\pi)+Ce^{-\Delta \tau}\sin (\tau - \frac{1}{4} \pi
)}{e^{\Delta \tau}\cos (\tau +\frac{1}{4} \pi)+Ce^{-\Delta \tau}
\cos (\tau -\frac{1}{4} \pi)}.
\end{equation}
Therefore, the value of $\mathcal{P} =\Psi \Psi ^{\ast}$ may be
represented as (see \cite{Pipa})
\begin{equation}
\mathcal{P}(\chi ,\tau)\sim (\bar{\xi ^{2}})^{-1/2}e^{-\xi ^{2}/(2
\bar{\xi ^{2}})},
\end{equation}
where
\begin{equation}
\bar{\xi ^{2}}=\frac{1}{Re[4y]}=\frac{1}{2}\left(p^{2}\cos
^{2}\theta +\frac{1}{p^{2}}\sin ^{2}\theta \right),
\end{equation}
\[
p=e^{\frac{1}{4}\beta \tau}, \theta = \tau -\frac{1}{4}\pi.
\]
The interpretation of Eq. (19)-(20) is evident enough: the wave
packet, which initially was localized as a whole in left well
became vigorously oscillate so that the probability amplitude take
the non-zero values in right well. The transition coefficient may
be easily found as an integral of probability density $\mathcal{P}
(\xi ,\tau)$ between the classical turning points in the right
well averaged by infinite time interval
\begin{equation}
D=\int _{\xi _{l}}^{\xi _{r}}d\xi \left\langle \Psi
\Psi^{\ast}\right\rangle _{t\in (0,\infty)} ,
\end{equation}
where
\[
\xi _{l,r}=\sqrt{\eta ^{2} \pm\frac{\sqrt{2}}{2}\sigma
^{-3/2}\sqrt{\hbar \omega _{0}}}.
\]
The numerical calculations shows that penetrability coefficient
will periodically changes with a growthing of walls velocity  from
zero to value close to 1 for the different values of the velocity.
More over, the resonant velocities which determined by the
``classical'' condition of the parametrical resonance (7)
demonstrate resonant behavior in the considered situation too: the
magnitude of the penetrability coefficients reach maximal values
just for the same velocities.

In Table 1 we show for the example some values of the
penetrability coefficients calculated for the $YFeO_{3}$ ($\beta
=1\cdot 10^{-4}$).

In conclusion let's describe the general script of the
parametrical evolution of the surface waves in the moving domain
wall.

When the velocities of the wall are far from the resonance, the
wall remains flat and no vibrations occur. It can be explained by
the fact that ground state which corresponds to the flat wall are
separated by the width energy gap from the excited states which
corresponds to the wall with a surface waves.

But such a gap can be overcomed due to the parametrical tunneling,
which probability reaches the macroscopical values when the wall's
velocity close to one of its resonant values. After excitation of
the lowest mode it's following evolution will be determined by the
classical parametrical resonance, and due to the energy
conservation the growth of the surface wave's amplitude will leads
to the braking of the wall, and the peculiarities of the mobility
may occur likewise there were observed in experiments.
\begin{table}
\begin{center}
\caption{The probability of the resonant transitions in spectrum
of moving domain wall for the some resonant velocities}
\begin{tabular}{|l|l|l|}
\hline Numbers of resonance & Resonant velocities (cm/s) &
Probability D
\\ \hline $0$ & $\allowbreak 2.0\times 10^{6}$ & $1$ \\ \hline $1$
& $\allowbreak 6.\,\allowbreak 066\,3\times 10^{5}$ & $0.934$ \\
\hline $3$ & $2.\,\allowbreak 110\,2\times 10^{5}$ & $0.865$ \\
\hline $5$ & $1.\,\allowbreak 270\,7\times 10^{5}$ & $0.649$ \\
\hline $10$ & $\allowbreak 63630.0$ & $0.501$ \\ \hline
\end{tabular}
\end{center}
\end{table}

\end{document}